# Fourier Analysis of Redshift Space Distortions and the Determination of $\Omega$

Shaun Cole *Department of Physics, University of Durham, Science Laboratories, South Road, Durham DH1 3LE.*

Karl B. Fisher *Institute of Astronomy, Madingley Road, Cambridge CB3 OHA.*

and David H. Weinberg *Institute for Advanced Study, Olden Lane, Princeton, NJ 08540, USA.*

e-mail: SHAUN.COLE@DURHAM.AC.UK, FISHER@MAIL.AST.CAM.AC.UK, DHW@GUINNESS.IAS.EDU

**Summary.** The peculiar velocities of galaxies distort the pattern of galaxy clustering in redshift space, making the redshift space power spectrum anisotropic. In the linear regime of gravitational instability models, the strength of this distortion depends only on the ratio $\beta \equiv f(\Omega)/b \approx \Omega^{0.6}/b$, where $\Omega$ is the cosmological density parameter and $b$ is the bias parameter. We derive a linear theory estimator for $\beta$ based on the harmonic moments of the redshift space power spectrum. Using $N$-body simulations, we examine the impact of non-linear gravitational clustering on the power spectrum anisotropy and on our $\beta$-estimator. Non-linear effects can be important out to wavelengths $\lambda \sim 50 \mathrm{h}^{-1} \mathrm{Mpc}$ or larger; in most cases, they lower the quadrupole moment of the power spectrum and thereby depress the estimate of $\beta$ below the true value. With a sufficiently large redshift survey, the scaling of non-linear effects may allow separate determinations of $\Omega$ and $b$.

We describe a practical technique for measuring the anisotropy of the power spectrum from galaxy redshift surveys, and we test the technique on mock catalogues drawn from the $N$-body simulations. Preliminary application of our methods to the 1.2 Jy IRAS galaxy survey yields $\beta_{est} \sim 0.3 - 0.4$ at wavelengths $\lambda \sim 30 - 40 \mathrm{h}^{-1} \mathrm{Mpc}$. Non-linear effects remain important at these scales, so this estimate of $\beta$ is probably lower than the true value.

## 1 Introduction

A fundamental assumption of the standard cosmological model is that the universe has no preferred direction, so that galaxy clustering is, statistically speaking, isotropic. However, we usually map the distribution of galaxies in *redshift space*, using redshift rather than true distance as the radial coordinate. In redshift space, peculiar velocities displace galaxies along the line of sight and introduce a preferred direction in the pattern of galaxy clustering. One can exploit this clustering anisotropy to measure properties of the galaxy velocity field and constrain the value of $\Omega$, the cosmological density parameter (Sargent & Turner 1977). This paper describes some practical methods for carrying out such measurements, and it explores their limitations.

In the linear regime of gravitational instability models, the anisotropy of clustering takes a very simple form in Fourier space, as shown by Kaiser (1987; see also M$^{\mathrm{c}}$Gill 1990). The strength of an individual plane wave is amplified by a factor that depends on the angle between the wavevector and the line of sight,

$$\delta_{\mathbf{k}}^{S} = \delta_{\mathbf{k}}^{R} \left(1 + \beta \mu_{\mathbf{kl}}^{2}\right) , \quad (1.1)$$

where $\delta_{\mathbf{k}}^{R}$ and $\delta_{\mathbf{k}}^{S}$ denote the Fourier amplitudes in real and redshift space respectively, and $\mu_{\mathbf{kl}}$ is the cosine of the angle between the wavevector, $\mathbf{k}$, and the line of sight, $\mathbf{l}$.[1] The degree of amplification is controlled by $\beta \equiv f(\Omega)/b$, where $f(\Omega) \approx \Omega^{0.6}$ is the logarithmic derivative of the fluctuation growth rate (see Peebles 1980, §14), and $b$ is the "bias factor", an assumed constant of proportionality between galaxy and mass fluctuations. In linear theory, $\beta$ fixes the ratio of the peculiar velocity field to the gravitational acceleration field inferred from the galaxy distribution. Upon taking the ensemble average of the modulus of (1.1), one sees that the power spectrum in redshift space is anisotropic and given by

$$\begin{aligned}P^{S}(k, \mu_{\mathbf{kl}}) &\equiv \langle \delta_{\mathbf{k}}^{S} \delta_{\mathbf{k}}^{S*}\rangle \\ &= P^{R}(k)\left(1 + \beta \mu_{\mathbf{kl}}^{2}\right)^{2} ,\end{aligned} \quad (1.2)$$

where the real space power spectrum, $P^{R}(k)$, is assumed to be an isotropic function of $k$. Kaiser remarked that the angular dependence of the redshift space power spectrum could be used to measure the quantity $\beta$, and hence to constrain $\Omega$ if the bias parameter could be determined independently.

This paper extends Kaiser's analysis in several ways. First, we show that the expression (1.2) has a simple expansion in harmonic moments, and that the ratio of the quadrupole and monopole moments of $P^{S}(k,\mu)$ provides a simple estimator for $\beta$ in the linear regime. Second, we examine the effects of non-linear evolution on $P^{S}(k,\mu)$ using cosmological $N$-body simulations. Non-linearities can be important out to quite large scales, so they complicate the task of inferring $\beta$ from the power spectrum, but they also split the degeneracy between $\Omega$ and $b$ that exists in linear theory. Finally, we describe a practical method of measuring $P^{S}(k,\mu)$ from galaxy redshift surveys. Equation (1.2) assumes that the observer is at a great distance from the volume occupied by the survey, so that all the galaxies in the survey can be treated as though they lie along the same line of sight. We develop a way to handle this constraint, given a redshift survey of (fairly) arbitrary geometry, and we show how to correct the estimated power spectrum for the systematic effect of shot noise and for the effect of selecting galaxies within a finite window. We test our method by applying it to mock redshift surveys drawn from $N$-body simulations, and we present preliminary results from an analysis of the 1.2 Jy *IRAS* galaxy redshift survey (Fisher 1992; Fisher *et al.* 1993a; hereafter FDSYH).

Many previous studies of clustering anisotropy have employed the redshift space correlation function, measured as a function of separations parallel and perpendicular to the line of sight. On small scales, the random velocities of galaxies in virialized clusters cause the well known "finger-of-God" effect, which stretches clusters along the line of sight in redshift space. On large scales, inflow compresses overdense regions along the line of sight, and outflow expands underdense regions. The "finger-of-God" effect stretches contours of the correlation function along the line of sight at small separations, while streaming motions compress the contours at large separations. Davis & Peebles (1983) and Bean *et al.* (1983) used the small scale distortion to estimate the pairwise velocity dispersion of galaxies. The predicted form of the distortion on large scales was first calculated by Lilje & Efstathiou (1989), and later by M$^{\rm c}$Gill (1990), and Hamilton (1992). Hamilton (1993a) has estimated $\beta$ from the large scale anisotropy of the correlation function, using the *IRAS* 2 Jy survey of Strauss *et al.* (1992). Fisher *et al.* (1993b) have measured the redshift space correlation function of the *IRAS* 1.2 Jy survey and examined both small scale and large scale anisotropies.

The correlation function and the power spectrum form a Fourier conjugate pair, so complete knowledge of one is equivalent to complete knowledge of the other. However, estimators of the two quantities may behave quite differently when applied to finite and noisy data sets. There are several advantages to characterizing large scale clustering in terms of the power spectrum instead of the correlation function. First, uncertainty in the mean galaxy density affects only the $k = 0$ mode of the power spectrum, while standard techniques for measuring the correlation function suffer random and systematic errors at all scales if the mean density is not known *a priori* (see Hamilton [1993b] for partial solutions to these problems). Second, the error properties of the power spectrum are cleaner than those of the correlation function, which greatly simplifies the quantitative testing of models. In particular, if primordial fluctuations are Gaussian, then Fourier modes of the density field are statistically independent in the linear regime, though one must take account of correlations introduced by estimating the power spectrum from a finite data sample. Third, theoretical studies suggest that the transition between Fourier modes that are evolving according to linear theory and shorter wavelength modes that have gone non-linear is fairly sharp (e.g. Little, Weinberg & Park 1991). Fourier transforming the spectrum mixes different physical scales, complicating the transition between the linear and non-linear regimes and hence the comparison between theory and observation. Finally, the power spectrum is more immediately related to the predictions of theoretical models, and on large scales the mapping from real space to redshift space is most naturally described in the Fourier regime.

---

[1] Throughout this paper we adopt the notation $\mu_{\mathbf{x}_1 \mathbf{x}_2}$ as the cosine of the angle between $\mathbf{x}_1$ and $\mathbf{x}_2$, and we use the superscripts $R$ and $S$ to denote quantities measured in real and redshift space respectively.

A number of authors have estimated $P^s(k)$ directly from galaxy redshift catalogues (Baumgart & Fry 1991; Gramman & Einasto 1991; Peacock 1991; Peacock & Nicholson 1991; Vogeley *et al.* 1992; Park, Gott & da Costa 1992; FDSYH; Feldman, Kaiser & Peacock 1993). The analysis in this paper overlaps most with the studies of the *IRAS* redshift surveys by FDSYH and Feldman *et al.* (1993), and with the $N$-body study of Gramman, Cen & Bahcall (1993). However, those papers examine only the spherically averaged power spectrum — the monopole moment of $P^s(k, \mu)$ — and they concentrate on the shape of the spectrum as a function of $k$. We focus instead on determining $\Omega$ and $b$ from the angular anisotropy of $P^s(k, \mu)$. We also undertake a more systematic exploration of non-linear effects.

Section 2 discusses theoretical aspects of the redshift space power spectrum. In §2.1 we derive a linear theory estimator for $\beta$ based on harmonic moments of the power spectrum. In §2.2 we explore non-linear effects using cosmological $N$-body simulations. In §2.3 we relate our analysis to the correlation function analysis of Hamilton (1992). Section 3 presents a practical method for measuring the redshift space power spectrum and estimating $\beta$ from galaxy redshift surveys. In §4 we apply this method to the 1.2 Jy *IRAS* redshift survey. Section 5 summarizes our conclusions. The two appendices contain mathematical details of the calculations in §2.3 and §3. Throughout the paper, we express distances in $h^{-1}$ Mpc, where $h \equiv H_0/(100 \, \text{km s}^{-1} \, \text{Mpc}^{-1})$.

## 2 Theory

### 2.1 A LINEAR THEORY ESTIMATOR FOR $\beta$

It is convenient and useful to characterize the redshift space power spectrum in terms of multipole moments. These moments define the decomposition of $P^s(k, \mu)$ into Legendre polynomials, hereafter denoted $\mathcal{P}_l(\mu)$,

$$P^s(k, \mu) = \sum_{l=0}^{\infty} P_l^s(k) \mathcal{P}_l(\mu) . \qquad (2.1)$$

The multipole moments, $P_l^s(k)$, can be computed by the inversion formula

$$P_l^s(k) \equiv \frac{2l+1}{2} \int_{-1}^{+1} d\mu P^s(k, \mu) \mathcal{P}_l(\mu) . \qquad (2.2)$$

For reference, the first three even Legendre polynomials are $\mathcal{P}_0(\mu) = 1$, $\mathcal{P}_2(\mu) = (3\mu^2 - 1)/2$, and $\mathcal{P}_4(\mu) = (35\mu^4 - 30\mu^2 + 3)/8$.

In linear theory, $P^s(k,\mu)$ has the simple angular dependence of equation (1.2), which is characterized completely by its monopole, quadrupole, and hexadecapole moments. Equation (1.2) contains no odd powers of $\mu$ and no powers higher than $\mu^4$, so all odd multipoles and all moments with $l > 4$ must vanish according to equation (2.1). The non-vanishing moments can be obtained easily by direct integration:

$$P_0^s(k) = \left(1 + \frac{2}{3}\beta + \frac{1}{5}\beta^2\right) P^R(k)$$

$$P_2^s(k) = \left(\frac{4}{3}\beta + \frac{4}{7}\beta^2\right) P^R(k) \qquad (2.3)$$

$$P_4^s(k) = \left(\frac{8}{35}\beta^2\right) P^R(k)$$

$$P_l^s(k) = 0 \text{ for } l \neq 0, 2, \text{ or } 4.$$

As we will see in §2.3, this behaviour is analogous to that of the redshift space correlation function described by Hamilton (1992). When clustering is non-linear, one can still perform a multipole decomposition of the power spectrum as in equation (2.1), but in this regime one expects that $l > 4$ moments will contribute and that the low-order moments will depart from the values in equation (2.3). In particular, the "finger-of-God" distortion reverses the sign of the quadrupole term on small scales because it weakens fluctuations along the line of sight instead of amplifying them.

If we have measured $P^s(k,\mu)$ on large scales and decomposed it into multipole moments, then we can estimate $\beta$ from the ratio of the quadrupole term to the monopole term. The linear theory prediction, which follows directly from (2.3), is

$$\frac{P_2^s(k)}{P_0^s(k)} = G(\beta) \equiv \frac{\frac{4}{3}\beta + \frac{4}{7}\beta^2}{1 + \frac{2}{3}\beta + \frac{1}{5}\beta^2}. \qquad (2.4)$$

This function is well behaved over the interesting range of $\beta$, growing almost linearly from $G \approx 0.13$ at $\beta = 0.1$ to $G \approx 1.02$ at $\beta = 1$, so the inversion from $G(\beta)$ to $\beta$ is stable and straightforward. In principle one can also obtain $\beta$ from the ratio $P_4^s(k)/P_0^s(k)$, but in practice we find that the hexadecapole is too sensitive to noise and non-linearities in $P^s(k,\mu)$ to yield useful estimates.

One could devise other schemes for extracting $\beta$ from measurements of $P^s(k,\mu)$, but there are two advantages to the estimator defined by equation (2.4). First, the monopole and quadrupole moments provide a "low-order" description of the anisotropy predicted by equation (1.2), so an estimator based on them should be relatively insensitive to noise. Second, the power spectrum estimator described in §3 is most naturally expressed in terms of multipoles, because the correction for a finite data sample is described most simply in terms of the correction to each multipole moment. It therefore makes sense to base the $\beta$ estimator directly on these multipole moments.

The techniques described by FDSYH and Feldman *et al.* (1993) yield estimates of the direction-averaged redshift space power spectrum, i.e. the monopole component of $P^s(k,\mu)$. One can then derive $\beta$ by measuring the amplification of the redshift space power spectrum over the real space power spectrum, but only if the real space spectrum has itself been estimated by some other method (e.g. from an angular catalog). Our approach, on the other hand, exploits the angular dependence of the redshift space spectrum at fixed $k$, and the real space power spectrum cancels out of the estimator (2.4).

## 2.2 NON-LINEAR EFFECTS

In order to understand how non-linearities can affect $P^s(k,\mu)$ and the $\beta$ estimator defined by equation (2.4), it is helpful to recap Kaiser's (1987) linear theory derivation of the power spectrum anisotropy, equation (1.2). For simplicity, we will assume a volume limited galaxy sample, so that terms involving derivatives of the selection function vanish. If the small scale velocity dispersion can be neglected on the scale of interest, then the galaxy density in redshift space is related to the galaxy density in real space by the Jacobian of the transformation between the two coordinate systems,

$$\rho^{S}(\mathbf{s}) = \rho^{R}(\mathbf{r}) \left[1 + \frac{U(\mathbf{r}) - U(\mathbf{0})}{r}\right]^{-2} \left[1 + \frac{dU(\mathbf{r})}{dr}\right]^{-1} , \qquad (2.5)$$

where $U(\mathbf{r})$ is the radial component of the peculiar velocity. The first multiplying factor describes the effect of the peculiar velocity on the cross sectional area of an element of fixed angular size, while the second factor describes compression or expansion along the line of sight. The peculiar velocity associated with a plane wave of wavenumber $k$ and amplitude $\Delta$ is $U \sim \Delta/k$, while $dU/dr \sim kU \sim \Delta$. We can therefore set the first factor to unity for a distant observer at $r \gg 1/k$, obtaining

$$\rho^{S}(\mathbf{s}) = \rho^{R}(\mathbf{r}) \left(1 + \frac{dU}{dr}\right)^{-1} . \qquad (2.6)$$

If $|dU/dr| \ll 1$, then an expansion of (2.6) yields

$$\rho^{S}(\mathbf{s}) = \rho^{R}(\mathbf{r}) \left(1 - \frac{dU}{dr}\right) , \qquad (2.7)$$

$$\Longrightarrow \quad \delta^{S}(\mathbf{s}) \equiv \rho^{S}(\mathbf{s})/\bar{\rho} - 1 = \delta^{R}(\mathbf{r}) - \frac{dU}{dr} - \delta^{R}(\mathbf{r})\frac{dU}{dr} . \qquad (2.8)$$

Now consider a single Fourier mode of the galaxy density field, with wavevector $\mathbf{k}$ and modulus $\Delta_{\mathbf{k}}$. The associated real-space density perturbation is

$$\delta^{R}(\mathbf{r}) = \Delta_{\mathbf{k}} \cos(\mathbf{k} \cdot \mathbf{r} + \varphi) , \qquad (2.9)$$

and in linear theory the corresponding velocity perturbation is

$$\mathbf{v}(\mathbf{r}) = -\beta \frac{\mathbf{k}}{k^{2}} \Delta_{\mathbf{k}} \sin(\mathbf{k} \cdot \mathbf{r} + \varphi) . \qquad (2.10)$$

In equation (2.10) we have assumed a linear bias between the galaxy and mass perturbations, $\Delta_{\mathbf{k}} = b \Delta_{\mathbf{k}}^{mass}$. The radial component of the velocity perturbation is

$$U(\mathbf{r}) = \mathbf{v}(\mathbf{r}) \cdot \hat{\mathbf{r}} = -\frac{\mu_{\mathbf{kl}}\beta}{k} \Delta_{\mathbf{k}} \sin(\mathbf{k} \cdot \mathbf{r} + \varphi) , \qquad (2.11)$$

$$\Longrightarrow \quad \frac{dU}{dr} = -\mu_{\mathbf{kl}}^{2}\beta\Delta_{\mathbf{k}}\cos(\mathbf{k} \cdot \mathbf{r} + \varphi) = -\mu_{\mathbf{kl}}^{2}\beta\delta^{R}(\mathbf{r}) . \qquad (2.12)$$

Combining this result with equation (2.8) yields

$$\delta^{S}(\mathbf{s}) = \delta^{R}(\mathbf{r})[1 + \mu_{\mathbf{kl}}^{2}\beta + \mu_{\mathbf{kl}}^{2}\beta\delta^{R}(\mathbf{r})] \qquad (2.13)$$

$$= \delta^{R}(\mathbf{r})(1 + \mu_{\mathbf{kl}}^{2}\beta), \quad \text{if} \quad |\delta^{R}(\mathbf{r})| \ll 1 . \qquad (2.14)$$

The last result is simply equation (1.1), and equation (1.2) for the power spectrum follows upon taking the appropriate ensemble average.

Four distinct guises of the linear approximation appear in this derivation, the assumptions that:

(1) the small scale velocity dispersion can be neglected,

(2) $|dU/dr| \ll 1$, so that the Jacobian can be expanded as it is in equation (2.7),

(3) the velocity and density perturbations satisfy the linear theory relation (2.10),

(4) $|\delta^{R}(\mathbf{r})| \ll 1$, so that equation (2.14) follows from equation (2.13).

We have also made the distant observer approximation to obtain equation (2.6), but this is a geometrical approximation that is not specific to linear theory. Assumptions (1)–(4) all hold on scales that are truly in the linear regime. However, it is helpful to distinguish them because they break down at different rates as one moves to smaller scales. Furthermore, the relative severity of the breakdowns will vary from model to model, since the accuracy of assumption (1) depends on the amplitude of small scale dispersions, (2) on the magnitude of large scale velocity gradients, (3) on the contrast of mass density fluctuations, and (4) on the contrast of galaxy density fluctuations. The multiplicity of non-linear effects makes their combined behaviour complicated, but it also offers the hope of achieving separate constraints on model parameters that would be degenerate in linear theory. For instance, in a model with very low $\Omega$, one would expect violations of (1) and (2) to be small relative to violations of (3) and (4).

In what follows, we will refer to violations of assumption (1) as "dispersion non-linearity," violations of (2) as "gradient non-linearity," violations of (3) as "dynamical non-linearity," and violations of (4) as "contrast non-linearity." The boundary between dispersion and gradient non-linearity is not a clear one, since the very existence of dispersion in the velocity field contradicts the notion of a simple Jacobian transformation between real space and redshift space. Nonetheless, it is useful to maintain some conceptual distinction between a form of non-linearity that derives primarily from virialized objects and a form that reflects the amplitude of larger scale velocity gradients. Operationally, we can regard dispersion non-linearity as the component that is removed by smoothing the galaxy velocity distribution on small scales (a few Mpc). It is difficult to calculate the impact of these various non-linear effects analytically, even if they are treated in isolation. For now, we will limit ourselves to a preliminary investigation based on cosmological $N$-body simulations.

The $N$-body simulations that we will analyze are a subset of those used by Little & Weinberg (1993) in their study of the void probability function. Each simulation has Gaussian initial conditions with the parametrized power spectrum of Efstathiou, Bond & White (1992). With Efstathiou *et al.*'s scale parameter $\Gamma$ set to 0.5, this power spectrum is a good fit to that of the standard cold dark matter model with $\Omega = 1$ and $h = 0.5$. We adopt a lower value of the scale parameter, $\Gamma = 0.25$, to obtain a spectral shape closer to that implied by recent measurements of the power spectra and correlation functions of galaxies and galaxy clusters (Maddox *et al.* 1990; Dalton *et al.* 1992; Peacock & Nicholson 1991; Vogeley *et al.* 1992; FDSYH; Feldman *et al.* 1993). Each simulation occupies a periodic cube $300h^{-1}$Mpc on a side, and there are two independent realizations of each model. We evolve the simulations into the non-linear regime using a staggered-mesh PM code written by Changbom Park (Park 1990), with $100^3$ particles on a $200^3$ density-potential mesh. In this paper we present results from models with $\Omega = 1$ and $\Omega = 0.3$, bias factors $b = 1$ and 2, and cosmological constant $\Lambda = 0$. We have also examined models with $\Omega = 0.1$ and with $b = 3$. Initial power spectra are normalized so that the rms amplitude of mass fluctuations in spheres of radius $8h^{-1}$Mpc would be $\sigma_8 = 1/b$ if the initial conditions were extrapolated to $z = 0$ by linear theory. For the unbiased ($b = 1$) models we select a "galaxy" population randomly from the $N$-body particles. For the biased models, we select a biased subset of the particles according to the high peaks algorithm (Bardeen *et al.* 1986), so that for galaxies $\sigma_8^g = b\sigma_8 = 1$. More details of the simulations and the biasing procedure appear in Little & Weinberg (1993). From our point of view, the important feature of these simulations is their large physical volume, which allows us to measure the power spectrum with reasonable accuracy on large scales. However, the force resolution ($1.5h^{-1}$Mpc grid scale) is rather low, which probably causes us to underestimate the magnitude of small scale velocity dispersions. The simulations are adequate for the exploratory purposes of this paper, but higher resolution simulations would be desirable for a more detailed examination of velocity dispersion effects.

The light solid lines in Figure 1 show contours of the linear theory power spectrum (1.2) for $\beta = 1/2$ and $P^R(k)$ equal to the $\Gamma = 0.25$ power spectrum of our initial conditions. The axes $k_\parallel$ and $k_\perp$ indicate wavenumbers parallel and perpendicular to the line of sight, so $\mu_{\bf kl} = k_\parallel/(k_\parallel^2 + k_\perp^2)^{1/2}$. Note that the outer contours correspond to smaller spatial scales (higher $k$), and that the contour levels increase inwards. In linear theory, coherent flows amplify fluctuations with wavevectors close to the line of sight, so contours of the power spectrum are elongated in the $k_\parallel$ direction. If we imagine revolving these contours around the $k_\parallel$ axis (since there is a single line of sight direction but a full $2\pi$ range of tangential directions), we can describe this elongation as a prolate distortion of the spherically symmetric, real space power spectrum.

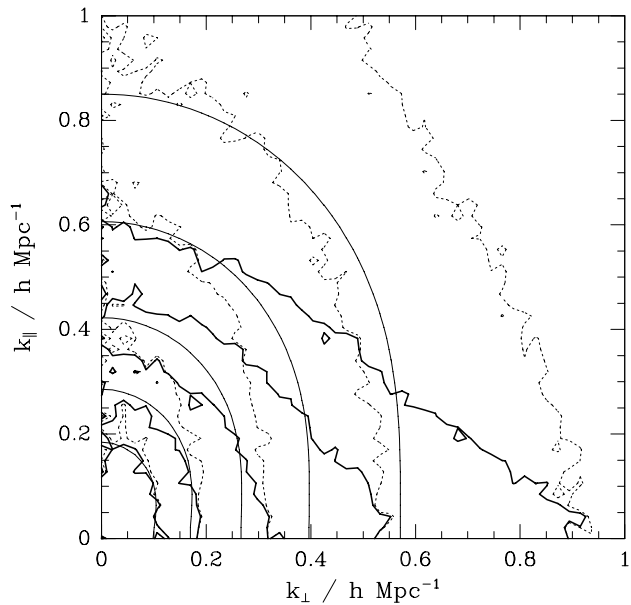

**Figure 1.** Contours of the redshift space power spectrum $P^S(k,\mu)$ in the plane defined by $k_\parallel$ and $k_\perp$, the components of the wavevector parallel and perpendicular to the line of sight — $\mu = k_\parallel/(k_\parallel^2 + k_\perp^2)^{1/2}$. Note that the *innermost* contours correspond to the *largest* physical scales. Contour levels increase from $P^S(k,\mu) = 2$ at the outside to $P^S(k,\mu) = 32$ at the inside. Light solid contours show the linear theory power spectrum, $P^S(k,\mu) = (1 + \beta\mu^2)^2 P^R(k)$, for $\beta = 1/2$ and the $\Gamma = 0.25$ real space spectrum $P^R(k)$. Heavy solid contours show $P^S(k,\mu)$ measured from an $N$-body simulation with $\Omega = 1$ and $b = 2$; non-linear effects are substantial for all but the innermost contour. Dotted contours show the power spectrum of the same simulation after the particle velocities have been smoothed with a spherical filter of radius $5h^{-1}$ Mpc. Smoothing removes "fingers-of-God" and restores the contour excursions to large $k_\parallel$.

The heavy contours in Figure 1 show the power spectrum measured from one of our $\Omega = 1$, $b = 2$ simulations. We shift the biased particles into "redshift" space, treating the the $z$−component of the peculiar velocity as the radial velocity, and we compute the power spectrum by Fast Fourier Transform (FFT). By identifying the $z$−axis with the line of sight, we automatically satisfy the distant observer constraint, and we ensure that the radial velocity field is periodic in the simulation volume, so that the boundary conditions assumed by the FFT are appropriate. Of course we cannot measure power spectra in the real universe this way; we leave that problem to §3. At the largest scale (innermost contour), the simulation power spectrum lies fairly close to the linear theory prediction, but at smaller scales there are substantial departures. In particular, the outer contours of the simulation power spectrum are oblate rather than prolate.

The oblate shape of these outer contours is a signature of dispersion non-linearity, which washes out small scale fluctuations along the line of sight. The reduction in small scale power compresses contours along the $k_\parallel$ axis. Oblate outer contours of $P^s(k_\perp, k_\parallel)$ are the Fourier analogue of prolate inner contours of the correlation function $\xi^s(r_p, \pi)$, which are also caused by "fingers-of-God" (see Fisher *et al.* 1993b). The dotted contours of Figure 1 show the power spectrum of the same simulation, except that we have smoothed the velocity field on small scales by replacing the velocity of each particle with the mean velocity of all particles within a $5h^{-1}$Mpc sphere centred upon it. This smoothing procedure removes the velocity dispersion of virialized clusters and groups, but it leaves the large scale flow intact. Smoothing alters the shapes of the outer contours dramatically, restoring their extensions to large $k_\parallel$. Comparison of dotted and solid curves shows that dispersion non-linearity significantly distorts the shapes of all but the innermost contour. The two sets of contours always agree when $k_\parallel/k_\perp \to 0$ because peculiar velocities do not alter the power spectrum at $\mu_{\bf kl} = 0$.

By squashing contours towards small $k_\parallel$ (small $\mu_{\bf kl}$), dispersion non-linearity suppresses the quadrupole moment of $P^s(k,\mu)$ relative to its monopole moment. It therefore depresses the value of $\beta$ inferred by our linear theory estimator (2.4). Figure 2a illustrates this effect for the $\Omega = 1$, $b = 2$ model. Solid lines show the ratio of the estimated $\beta$ to the true value, $\beta_{\rm true} = 1/b = 0.5$, as a function of wavelength $\lambda = 2\pi/k$. The two lines represent our two independent simulations;

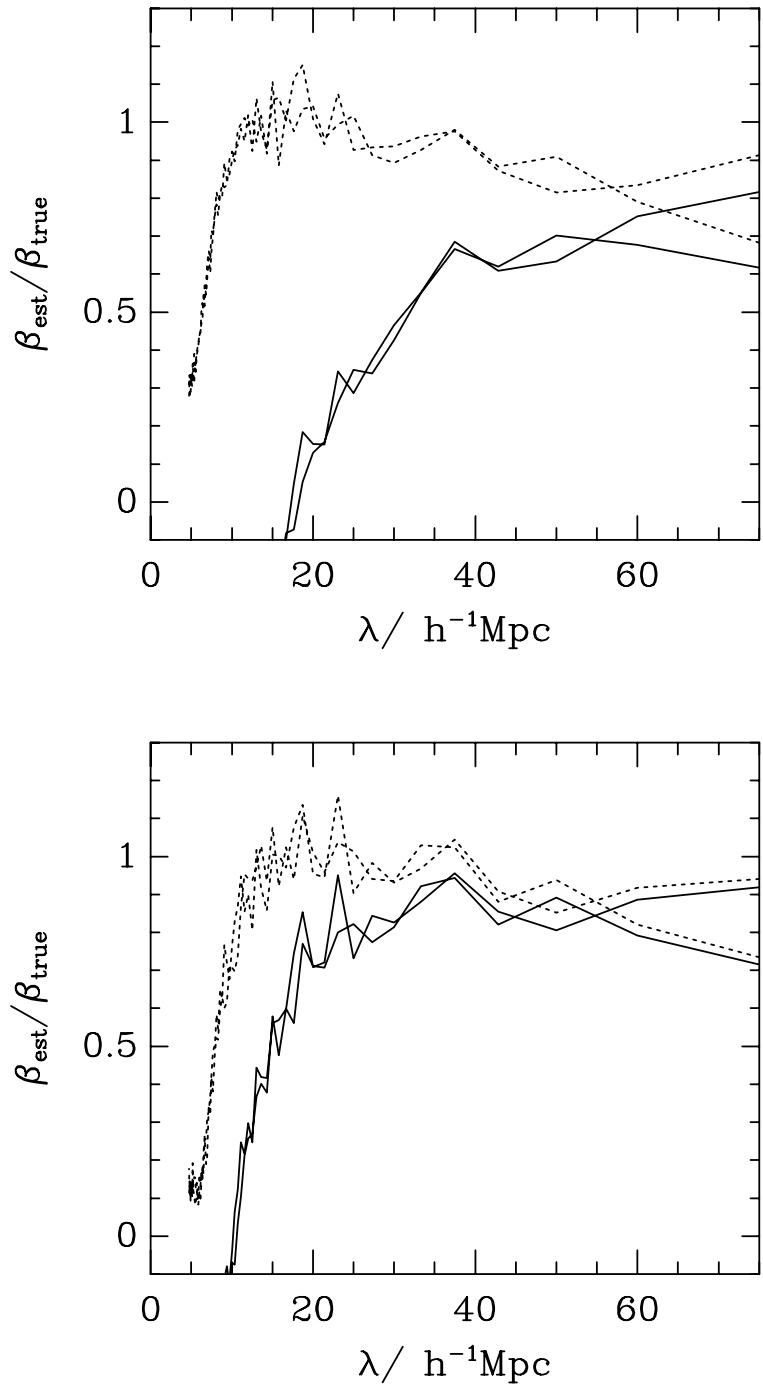

**Figure 2.** Effect of dispersion non-linearity on the linear theory $\beta$ estimator (2.4). The ratio of the estimated $\beta$ to the true value is plotted as a function of wavelength $\lambda \equiv 2\pi/k$. *a*) A model with $\Omega = 1$ and $b = 2$. Solid curves show results for two simulations without velocity smoothing; dispersion non-linearity depresses the estimated $\beta$ below the true value out to large scales. Dotted curves show results after smoothing the particle velocities over $5h^{-1}$Mpc; the estimated $\beta$ rises rapidly to the true value. *b*) Same as *a*), but for simulations with $\Omega = 0.3$ and $b = 2$. Smoothing has less effect in this model because velocity dispersions are smaller.

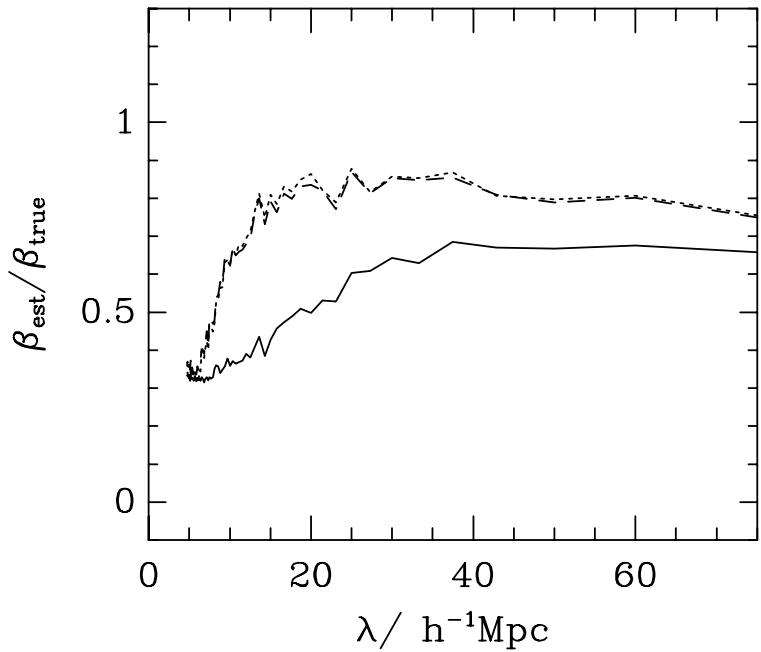

**Figure 3.** Effect of gradient non-linearity on the $\beta$ estimator (2.4). Solid and dotted lines show results for unbiased ($b = 1$) models with $\Omega = 1$ and $\Omega = 0.3$, respectively. Particle velocities have been smoothed over $5h^{-1}$Mpc in each simulation, and values plotted are the average results from our two independent runs of each model. The difference between the solid and dotted curves reflects the higher amplitude of the large scale velocity field in the $\Omega = 1$ model. The dashed line, nearly coincident with the dotted line, is estimated from the $\Omega = 1$ model after scaling all of its particle velocities (and the value of $\beta_{\rm true}$) by a factor of $0.3^{0.6}$, to match its velocity amplitude to that of the $\Omega = 0.3$ model.

the close match between them indicates that sample-to-sample fluctuations in $P^s(k,\mu)$ are small for wavelengths $\lambda \lesssim 60h^{-1}$Mpc, in volumes the size of our $300h^{-1}$Mpc simulation cubes. The dotted lines in this figure show the estimated values of $\beta/\beta_{\rm true}$ for the simulations after smoothing the velocity field. These curves rise much more quickly, leveling out at $\beta_{\rm est}/\beta_{\rm true} \approx 0.9$ by $\lambda = 40h^{-1}$Mpc. Comparison of the solid and dotted lines shows that dispersion non-linearity has a significant impact in this model even at $\lambda = 60h^{-1}$Mpc.

Figure 2b makes the same comparison for an $\Omega = 0.3$, $b = 2$ model. Once again the curves for the velocity-smoothed simulations rise more rapidly to their asymptotic value, but the effect of velocity dispersions is much smaller in this low-$\Omega$ model, as expected. Observational estimates of the galaxy pairwise velocity dispersion yield $\sigma_v \sim 350\,{\rm km\,s^{-1}}$ at separations $\sim 1h^{-1}$Mpc (Davis & Peebles 1983; Fisher *et al.* 1993b; other estimates range from $\sim 250\,{\rm km\,s^{-1}}$ by Bean *et al.* 1983 to $\sim 450\,{\rm km\,s^{-1}}$ by de Lapparent, Geller & Huchra 1988). The pairwise dispersion of the $\Omega = 1$, $b = 2$ simulations is $600\,{\rm km\,s^{-1}}$, and that of the $\Omega = 0.3$, $b = 2$ simulations is $250\,{\rm km\,s^{-1}}$, so the impact of dispersion non-linearity in the real universe is probably smaller than that indicated by Figure 2a but larger than that of Figure 2b.

The solid line in Figure 3 shows the $\beta$ inferred by our quadrupole estimator for the unbiased ($b = 1$), $\Omega = 1$ model, averaged over the two simulations. We have smoothed the velocity field over $5h^{-1}$Mpc as before, to remove the effects of dispersion non-linearity. Nonetheless, the $\beta_{\rm est}$ vs. $\lambda$ curve rises slowly, and it levels out at only $\beta_{\rm est} \sim 0.7\beta_{\rm true}$. The dotted line shows the corresponding estimate for the unbiased, $\Omega = 0.3$ model. Here the $\beta$-estimator performs much better, rising to a value $\beta_{\rm est} \sim 0.85\beta_{\rm true}$ by $\lambda = 20h^{-1}$Mpc, then leveling off. We have removed the small scale velocity dispersion by smoothing, and the large scale spatial clustering in the two models is virtually identical because they have the same initial power spectrum. The difference in results must therefore reflect the difference in the amplitude of the large scale velocity field, which influences $\beta_{\rm est}$ through gradient non-linearity. To confirm this point, we scale down the individual particle velocities of the $\Omega = 1$ model by a factor $0.3^{0.6}$, then smooth the velocity field, measure the power spectrum, and estimate $\beta$ as before. We obtain the dashed line of Figure 3, which is virtually indistinguishable from the dotted line that represents the $\Omega = 0.3$ model. The detailed match of individual features in these curves arises because we use the same initial random phases for the $\Omega = 1$ and $\Omega = 0.3$ simulations.

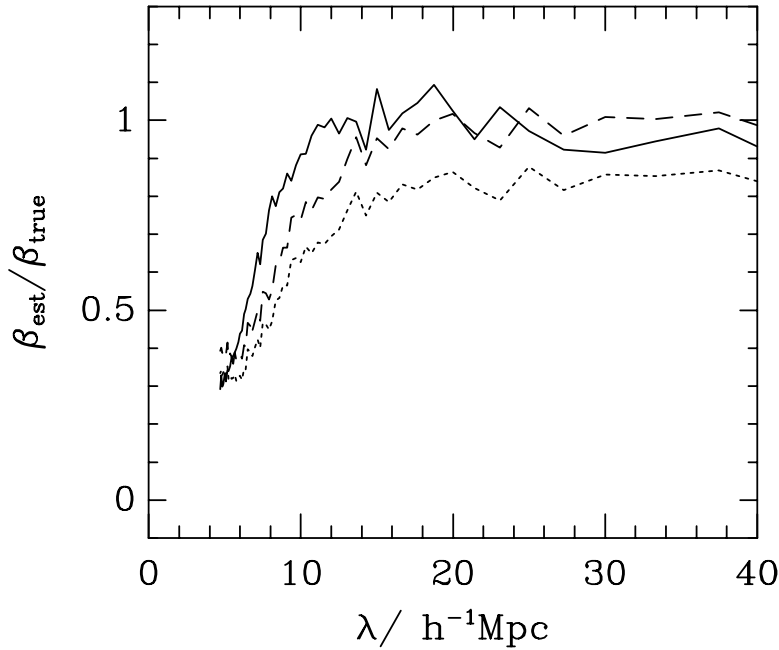

**Figure 4.** Effect of dynamical non-linearity on the $\beta$ estimator (2.4). The solid line represents the model with $\Omega = 1$ and $b = 2$ ($\beta = 0.5$). The dotted line represents the model with $\Omega = 0.3$ and $b = 1$ ($\beta = 0.49$). Both curves are averaged over two simulations, and particle velocities have been smoothed over $5h^{-1}$Mpc. The dashed line shows the dotted curve multiplied by a factor of $1/0.85$, so that it levels out at $\beta_{\rm est}/\beta_{\rm true} \approx 1$. Although true $\beta$ values are nearly identical in the two models, mass fluctuations are non-linear out to larger scales in the unbiased, low-$\Omega$ model, so the $\beta_{\rm est}(\lambda)$ curve levels out at larger wavelengths. Dynamical non-linearity offers a potential tool for obtaining separate constraints on $\Omega$ and $b$, given a sufficiently large data set.

Increasing $\beta$ amplifies both dispersion and gradient non-linearity. We have already remarked that there is no clean separation between these two forms, but since we have smoothed away the small scale dispersions for both of the models shown in Figure 3, it seems appropriate to ascribe the remaining difference between them to gradient non-linearity. In the unbiased, $\Omega = 1$ model, this effect significantly depresses the $\beta$ inferred by our quadrupole estimator out to quite large scales. Results for an unbiased, $\Omega = 0.1$ model are similar to those for the $\Omega = 0.3$ model in Figure 3, which suggests that the residual impact of gradient non-linearity is small once $\Omega \lesssim 0.3$ ($\beta \lesssim 0.5$), at least with this degree of velocity smoothing.

In the linear regime, distortions of the redshift space power spectrum depend on a specific combination of physical parameters, $\beta = \Omega^{0.6}/b$. Clearly we would like to know the values of $\Omega$ and $b$ individually. In particular, given the evidence that $\beta < 1$ from small scale clustering (e.g. cluster mass-to-light ratios or the cosmic virial theorem), we would like to be able to distinguish a low density universe in which galaxies trace mass from a critical density universe with biased galaxy formation. Can we exploit the scaling of non-linear effects to make such a distinction?

To make the question specific, suppose that we analyze a large redshift sample and find that our quadrupole technique yields a stable estimate of $\beta \approx 0.5$ on large scales. We wish to distinguish between two alternative hypotheses: an $\Omega = 0.3$ model with $b = 1$ (and $\beta = 0.3^{0.6} = 0.49$), and an $\Omega = 1$ model with $b = 2$. The amplitude of large scale flows in these two models is the same, and small scale velocity dispersions will be similar, so gradient and dispersion non-linearities cannot help us. The observed power spectrum directly determines the amplitude of galaxy fluctuations, and hence the importance of contrast non-linearity. However, the amplitude of *mass* fluctuations is different in the two models, with the unbiased, $\Omega = 0.3$ model having larger fractional mass fluctuations at a given physical scale. One might therefore expect that dynamical non-linearity would appear in the low-$\Omega$ model on a scale where the $\Omega = 1$ model still satisfies equation (2.10), the linear theory relation between velocity and density.

Figure 4 illustrates just this behaviour. The solid line shows the ratio $\beta_{\rm est}/\beta_{\rm true}$ for our $\Omega = 1$, $b = 2$ simulations, with the velocity field smoothed over $5h^{-1}$Mpc. It rises to $\beta_{\rm est} \approx \beta_{\rm true}$ by $\lambda \approx 12h^{-1}$Mpc, then levels off. The dotted line represents the $\Omega = 0.3$ model, and it does not level off until $\lambda \approx 20h^{-1}$Mpc. The "turnover wavelength" of the unbiased model is nearly double that of the biased model, reflecting the importance of dynamical non-linearity at larger scales. In the region where $\beta_{\rm est}$ is stable, $\lambda \approx 20 - 40h^{-1}$Mpc, the value for the $\Omega = 0.3$ model is only $\beta_{\rm est} \approx 0.85\beta_{\rm true}$. The difference from the $\Omega = 1$ model, which has $\beta_{\rm est} \approx \beta_{\rm true}$ in this region, presumably reflects the continuing influence of dynamical non-linearity, since other non-linear effects should be similar in the two models. With a sufficiently large observational sample, we would expect to recover $\beta_{\rm est} \approx \beta_{\rm true}$ at larger wavelengths. However, presented with the dotted curve and no *a priori* knowledge of $\Omega$, we might mistake the stable value found between 20 and $40h^{-1}$Mpc for $\beta_{\rm true}$, and rescale the height of the $\beta_{\rm est}/\beta_{\rm true}$ vs. $\lambda$ curve accordingly. The result would be the dashed curve of Figure 8, which is simply the dotted curve multiplied by $1/0.85$. Although this curve lies closer to the solid curve that represents the $\Omega = 1$ results, one can still discern the larger scale of non-linearity in the unbiased model.

The moral of this section is that non-linear effects are a nuisance, but a potentially useful nuisance. Non-linearity can affect the anisotropy of the redshift space power spectrum out to quite large scales — wavelengths of $60h^{-1}$Mpc or more — and the variety of effects prevents a simple analytic treatment. We have catalogued these effects and obtained some sense of their numerical importance, but we have not yet devised practical means for handling or exploiting them. Dispersion non-linearity seems likely to pose the thorniest problems. We have dealt with it here by smoothing the velocities in real space, but that approach is not possible with a redshift catalog. One strategy is to smooth the data in redshift space (equivalent to multiplying in the Fourier domain), perhaps using an anisotropic filter that takes advantage of the known anisotropy of "fingers-of-God." Another option is to estimate the pairwise velocity distribution on small scales from the redshift space correlation function, then use this distribution function to correct the power spectrum. Of course one can also compare the uncorrected data directly to theoretical models that include small scale dispersion, but here one confronts the problem that the pairwise velocity dispersion is a difficult quantity to compute reliably from theory (witness the continuing controversy over "velocity bias").

The dependence of dynamical non-linearity on mass fluctuations rather than galaxy fluctuations offers the tantalizing prospect of constraining $\Omega$ and the bias factor separately. Unfortunately, dispersion, gradient, and contrast non-linearities can mask these dynamical effects, making them difficult to exploit. Further analytic and numerical investigations should help this effort; a distinctive signature that can isolate dynamical non-linearity from other effects would be especially useful. The distinction between the unbiased, low-$\Omega$ model and the biased, $\Omega = 1$ model in Figure 4 is subtle, difficult to measure even in our $300h^{-1}$Mpc cubes of "perfect" data. From the current generation of galaxy redshift surveys, an estimate of $\beta$ is probably the most we should expect. However, huge redshift samples like the million-galaxy Sloan Digital Sky Survey (Knapp & Gunn 1993) will allow high precision measurements of the galaxy power spectrum, so separate determinations of $\Omega$ and $b$ should be possible.

2.3 COMPARISON TO CORRELATION FUNCTION ANALYSIS

It is informative to compare the method of estimating $\beta$ described in §2.1 to Hamilton's (1992; 1993a) method, which is based on the redshift space correlation function $\xi^s(s,\mu)$. Hamilton decomposes the redshift space correlation function into harmonic moments $\xi_l^s(s)$, defined by

$$\xi^s(s,\mu) = \sum_l \xi_l^s(s)\mathcal{P}_l(\mu) \ . \tag{2.15}$$

He then estimates the rational function $G(\beta)$ [equation (2.4)] from a rather elaborate combination of the quadrupole and monopole terms,

$$\hat{G}(\beta) = \frac{\xi_2^s(s)}{-\xi_0^s(s) + (3/s^3)\int_0^s \xi_0^s(s')s'^2 ds'} \ , \tag{2.16}$$

and derives $\beta$ accordingly.

In Appendix B, we show that the harmonic components of $\xi^s(k,\mu)$ can be expressed elegantly in terms of the harmonic components of $P^s(k,\mu)$. We give expressions for these components and for some linear combinations of $\xi_l^s(s)$, such as the denominator of (2.16), which are useful in comparing our formalism to that of Hamilton (1992). Here we make use of these identities to re-express (2.16) in terms of the $P_l^s(k)$. Substituting the identities (B.5) and (B.8) into (2.16), we obtain a simple and informative expression for Hamilton's estimator of $G(\beta)$,

$$\hat{G}(\beta) = \frac{\int_0^\infty dk k^2 P_2^s(k) j_2(ks)}{\int_0^\infty dk k^2 P_0^s(k) j_2(ks)} \ , \tag{2.17}$$

where $j_2(kr)$ is the usual spherical Bessel function. In the linear regime the ratio $P_2^s(k)/P_0^s(k)$ is constant and equal to $G(\beta)$, and equation (2.16) therefore yields an unbiased estimate of $\beta$. The effects of non-linearity enter at large $k$, where $P_2^s(k)$ is depressed or even changes sign and $P_0^s(k)$ is boosted. These effects depress the value of $\hat{G}(\beta)$ given by (2.16) at small $s$. Non-linear small scale power can affect $\hat{G}(\beta)$ out to quite large separations because of the oscillating weight function $j_2(ks)$. If the transition between linear and non-linear behaviour is sharp in the power spectrum, it will be fuzzy in the correlation function.

The power spectrum has relatively simple error properties on large scales, which give it some practical advantage over the correlation function. However, the crucial question is which quantity can be estimated more accurately on scales in the linear regime, and the answer to this question may be survey- and model-dependent.

## 3 Measuring $P^s(k,\mu)$ from a redshift survey

Kaiser's formula for the redshift space power spectrum, equation (1.2), is valid only if $\delta_{\mathbf{k}}^s$ is computed from data that subtend a small solid angle on the sky. Without this constraint the quantity $\mu_{\mathbf{kl}}$ is not even clearly defined, since there is no single line of sight direction l. Faced with the task of measuring $P^s(k,\mu)$ and estimating $\beta$ from a galaxy redshift survey, our approach is to first extract a subset of the redshift data such that all of the selected galaxies lie approximately along the same line of sight. Specifically, we select galaxies within a spherically symmetric window of radius $R_{sph}$ centred at a distance $L$ from the observer sufficient that $\theta \approx L/R_{sph} \ll 1$. Only galaxies that fall inside the window are used in estimating the power spectrum. However, we can repeatedly reposition the window within the volume defined by the geometry of the redshift survey and the requirement that $L \gg R_{sph}$, then average the results to obtain a final, combined estimate that utilizes most of the galaxies in the survey. A roving window of this form has practical advantages, since one can apply the same type of analysis to cone surveys or to "all sky surveys" with incomplete coverage near the galactic plane. In practice, the limited depth of existing redshift surveys forces one to compromise between using distant, poorly sampled data, which satisfy $\theta \ll 1$ but yield a noisy Fourier transform, or carrying out the transform with nearby, well sampled structure, for which the small-angle approximation is less accurate. At the end of this section we will calibrate the effect of a finite opening angle on the power spectrum estimate empirically, using a catalogue drawn from an $N$-body simulation.

Once the data are windowed to ensure an acceptably small opening angle, the estimates of Fourier modes at different $\mathbf{k}$ are no longer independent. Selecting galaxies within a window is equivalent to multiplying the galaxy density field by a window function (which is zero outside the windowed region). In Fourier space, this multiplication becomes a convolution with the Fourier transform of the window function, which introduces correlations between neighbouring modes over a range $\Delta k \approx 1/R_{sph}$. In any real application, one therefore expects the measured $P^s(k,\mu)$ to represent the true redshift space power spectrum *convolved* with a function related to the Fourier transform of the window used to define the sample. This convolution tends to make the measured redshift space power spectrum more isotropic, and therefore, if uncorrected, it causes a systematic *underestimate* of $\beta$. Fortunately, this effect can be calculated analytically (see below and Appendix A), and corrections can be applied to the estimated power spectrum.

Suppose that we wish to measure the power spectrum from a redshift survey characterized by mean galaxy density $\langle n \rangle$ and radial selection function $\phi(r)$ — i.e., in the absence of inhomogeneities the mean galaxy density at distance $r$ would be $\langle n \rangle \phi(r)$. A straightforward estimator for the Fourier mode amplitudes is the sum

$$S_{\mathbf{k}} = \frac{1}{\langle n \rangle V_w^{\frac{1}{2}}} \sum_{i=1}^{N} \frac{w(\mathbf{s}_i)}{\phi(\mathbf{s}_i)} e^{+i\mathbf{k}\cdot\mathbf{s}_i} \;, \qquad (3.1)$$

where $N$ is the total number of galaxies in the survey and $V_w \equiv \int d^3\mathbf{s}\, w(\mathbf{s})$ is the weighted volume of the sample. The summation in equation (3.1) is carried out in redshift space, and windowing of the data is introduced explictly through the window function $w(\mathbf{s})$. Again, we have in mind a window function that is non-zero over a region (e.g., a sphere) whose origin is displaced far enough from the central observer that it subtends a small angle. The simplest such function is a "top-hat" defined by centre $\mathbf{s}_c$ and radius $R_{sph}$: $w(\mathbf{s}) = \vartheta(|\mathbf{s} - \mathbf{s}_c|/R_{sph})$, where $\vartheta$ is a unit step function. The sharp edges of a top hat create nasty sidelobes in its Fourier transform, so in our analysis we adopt a "bowler-hat" window, defined as the convolution of a top hat with a Gaussian:

$$w(\mathbf{s}) = (2\pi)^{-3/2} R_g^{-3} \int d^3\mathbf{s}'\, \vartheta(|\mathbf{s}' - \mathbf{s}_c|/R_{sph})\, e^{-|\mathbf{s}-\mathbf{s}'|^2/2R_g^2}. \qquad (3.2)$$

Gaussian convolution apodizes the edges of the window and reduces the sidelobes in its Fourier transform. Results in this paper were computed using $R_g/R_{sph} = 0.1$.

We show in Appendix A that the ensemble average of the squared-modulus of $S_{\mathbf{k}}$ is

$$\langle S_{\mathbf{k}} S_{\mathbf{k}}^* \rangle = P^{sc}(k, \mu_{\mathbf{kl}}) + V_w |\tilde{w}(\mathbf{k})|^2 + \frac{1}{\langle n \rangle V_w} \int d^3\mathbf{s}\, \frac{|w(\mathbf{s})|^2}{\phi(s)} \;, \qquad (3.3)$$

where $\tilde{w}(\mathbf{k}) \equiv V_w^{-1} \int d^3\mathbf{s}\, w(\mathbf{s}) e^{+i\mathbf{k}\cdot\mathbf{s}}$ is the Fourier transform of window function, normalized to unity at $\mathbf{k} = 0$. The second term on the right hand side of (3.3) reflects the finite width of the window function in Fourier space; in the limit of a large volume, this term would contribute only at $\mathbf{k} = 0$. The last term represents the discreteness or "shot noise" contribution to the Fourier transform. The term of interest is $P^{sc}(k, \mu_{\mathbf{kl}})$, which is the convolution of the true redshift space power spectrum $P^s(k, \mu_{\mathbf{kl}})$ with the square of the window transform:

$$P^{sc}(k, \mu_{\mathbf{kl}}) \equiv \frac{V_w}{(2\pi)^3} \int d^3\mathbf{k}'\, P^s(k', \mu_{\mathbf{k'l}}) \left|\tilde{w}(\mathbf{k} - \mathbf{k}')\right|^2 \;. \qquad (3.4)$$

Multipole decomposition allows a simple treatment of the convolved redshift space power spectrum. If we define $P_l^{sc}(k)$ to be the multipole moments of the convolved spectrum $P^{sc}(k,\mu)$, analogous to the decomposition of $P^s(k,\mu)$ in equation (2.1), we can derive a relation between the true and convolved multipole moments (cf., Appendix A):

$$P_l^{sc}(k) = \int_0^\infty dk' k'^{2l} P_l^s(k') \mathcal{F}_l(k,k') ,$$
$$\equiv P_l^s(k) T_l^c(k) .$$
(3.5)

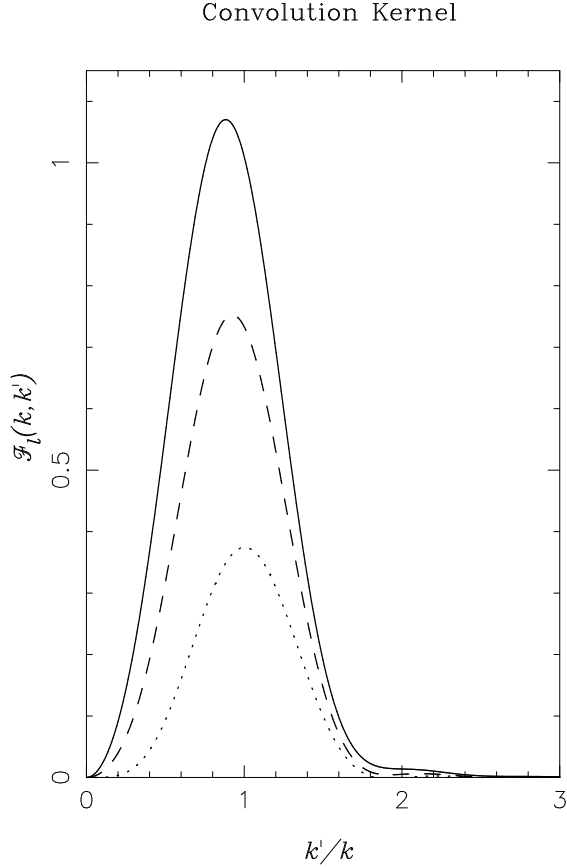

**Figure 5.** The convolution kernels $\mathcal{F}_l(k,k')$ that appear in the convolution of the multipole moments, equation (3.5), computed for our bowler-hat window function. The solid curve refers to the monopole moment ($l=0$), while the dashed and dotted curves refer to the quadrupole ($l=2$) and hexadecapole ($l=4$) moments, respectively.

Equation (A.15) of Appendix A defines the convolution kernel $\mathcal{F}_l(k,k')$. The "transfer" functions $T_l^c(k)$ describe the effect of the convolution on the multipole moments. In general, they depend on the wavenumber $k$, on the form of the power spectrum $P_l^s(k)$, and on the window geometry. However, for sensible choices of the window function $w(\mathbf{s})$, the convolution kernel $\mathcal{F}_l(k,k')$ is strongly peaked around $k=k'$, and in this case the integral that defines $T_l^c(k)$ depends only very weakly on $k$ and the power spectrum.

For a given value of $k$, we scale the radius of the window function $R_{sph}$ to give $\tilde{w}(k)=0$. The Fourier transform of a Gaussian is positive definite, so for the bowler-hat window the first zero occurs at the first zero of the top hat, which is $R_{sph}=4.4934/k$. This scaling has the advantage of making the second term in equation (3.3) identically zero. The only remaining term to be subtracted from $\langle S_\mathbf{k} S_\mathbf{k}^* \rangle$ is independent of $k$, so the shape of the estimated power spectrum (in a linear plot) is insensitive to uncertainties in the mean density $\langle n \rangle$ (FDSYH). This choice also simplifies the form of the transfer functions. The window transform $\tilde{w}(|\mathbf{k}-\mathbf{k}'|)$ is a function of the combination $|\mathbf{k}-\mathbf{k}'|R_{sph}$, so with $R_{sph}$ scaled to $k$ it becomes a function of $|\mathbf{k}-\mathbf{k}'|/k = [1+(k'/k)^2+2\mu_{\mathbf{kk}'}k'/k]^{1/2}$. Substituting into equation (A.15) and integrating over $\mu_{\mathbf{kk}'}$, we find

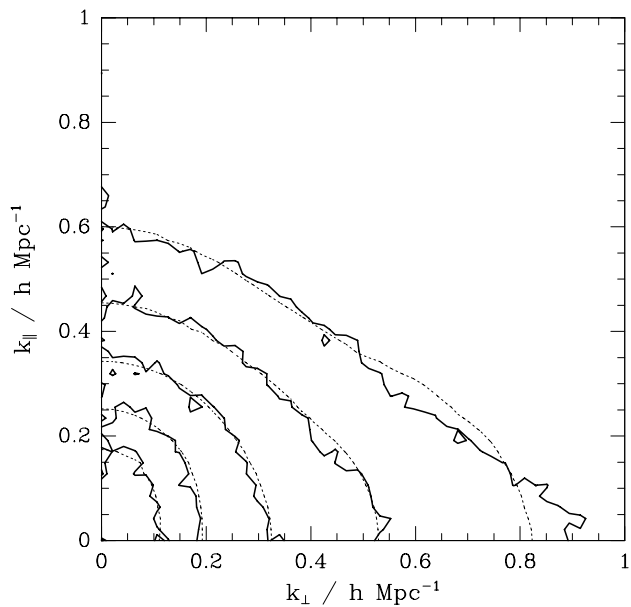

**Figure 6.** Recovery of the redshift space power spectrum by our "practical" algorithm. Format is similar to that of Figure 1. Heavy contours, repeated from that figure, show the redshift space power spectrum measured directly by FFT from an $\Omega = 1$, $b = 2$, $N$-body simulation. Dotted contours show the power spectrum recovered by the algorithm described in §3, which can also be applied to galaxy redshift surveys. These contours are reconstructed from just the monopole, quadrupole, and hexadecapole components. They agree well with contours of the FFT measurement at large scales. At small scales (outer contours), non-linearity introduces some power into the higher harmonics of $P^s(k,\mu)$, so the recovery is imperfect.

that the convolution kernels $\mathcal{F}_l(k,k')$ depend only on the ratio $k'/k$. If $P^s(k)$ is scale-free, then the functions $T_l^c(k)$ are independent of $k$.

Figure 5 shows the convolution kernels $\mathcal{F}_l(k,k')$ for our bowler-hat window, as functions of $k'/k$. The solid curve refers to the monopole moment ($l = 0$), while the dashed and dotted curves refer to the quadrupole ($l = 2$) and hexadecapole ($l = 4$) moments respectively. The multipole moments of the redshift space power spectrum measured in a finite window are the true multipole moments convolved according to equation (3.5) with the curves shown in Figure 5.

**Table 1**

| Spectral Index | $T_2^c/T_0^c$ | $T_4^c/T_0^c$ |
|---|---|---|
| $n = -2$ | 0.680 | 0.333 |
| $n = -1$ | 0.706 | 0.384 |
| $n = 0$ | 0.716 | 0.422 |
| $n = +1$ | 0.711 | 0.445 |

Table 1 gives the values of the transfer coefficients $T_l^c$ for our bowler-hat window, in the case where the true spectrum multipoles are $P_l^s(k) \propto k^n$. From Table 1, it is clear that the effect of convolution on the multipole moments of $P^s(k,\mu)$ is insensitive to the shape of the real space power spectrum. This is good news, since it means that one can correct the observed multipole moments of $P^{sc}(k,\mu)$ to recover the true redshift space power spectrum $P^s(k,\mu)$ without *a priori* knowledge of the power spectrum. However, Table 1 also shows that the convolution is a large effect for our choice of window function, so it cannot be neglected. For example, the ratio of the measured quadrupole to the measured monopole is suppressed by 30% from its true value.

We can test our "practical" method of estimating $P^s(k,\mu)$ by applying it to one of the $N$-body simulations described in §2.2. We first select a value of $k$ and define a bowler-hat window function with $R_{sph} = 4.4934/k$ and $R_g = 0.1 R_{sph}$. We view the simulation along a randomly chosen line of sight, **l**, and locate the bowler-hat window randomly within the simulation volume. For now, we define radial velocities as if we were viewing the data from an infinite distance, i.e. in the limit of a small opening angle. We carry out the Fourier sum $S_{\mathbf{k}}$, defined by equation (3.1), for a set of 50 randomly oriented $\mathbf{k}$-vectors. We average the squared-moduli $S_{\mathbf{k}} S_{\mathbf{k}}^*$ in bins of $\Delta\mu_{\mathbf{kl}} = 0.1$ and repeat the whole process $N_w$ times, choosing $N_w \gtrsim L_{\rm box}/R_{sph}^3$ to ensure that we adequately sample all the structure in the simulation. Subtracting the shot noise term of equation (3.3) leaves us with an estimate of $P^{s\,c}(k,\mu)$. We fit its angular dependence using a basis of Legendre polynomials $\mathcal{P}_l(\mu)$ and hence determine the coefficients $P_l^{s\,c}(k)$ in its multipole expansion [equation (A.8)]. To reconstruct the unconvolved $P^s(k,\mu)$ we need the multipole moments $P_l^s(k)$, which we obtain from the $P_l^{s\,c}(k)$ using equation (3.5) and the transfer coefficients $T_l^c(k)$ computed for our window function and the $\Gamma = 0.25$ power spectrum.

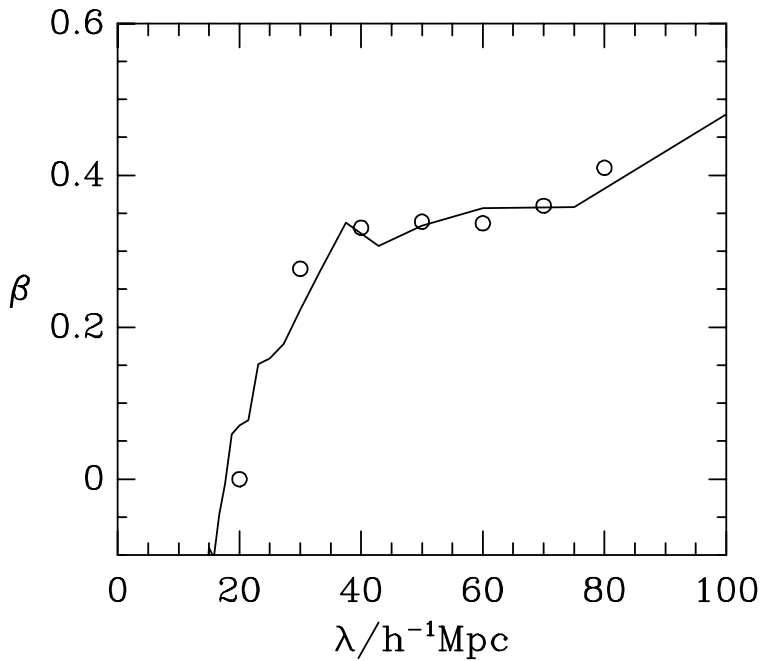

**Figure 7.** The value of $\beta$ inferred from the linear theory estimator (2.4). The solid line shows estimates based on the FFT measurement of $P^s(k,\mu)$ averaged over the two $\Omega = 1$, $b = 2$, $N$-body simulations. Circles show the result of applying our "practical" algorithm to the same simulations. Results of the two methods agree well, though the estimated $\beta$'s lie below the true value $\beta = 1/2$ because of the continuing effects of dispersion non-linearity on these scales (see Figure 2).

The heavy contours in Figure 6 show the redshift space power spectrum of the $\Omega = 1$, $b = 2$ $N$-body simulation, measured directly by FFT, just as in Figure 1. The dotted contours show the power spectrum measured as above, reconstructed from just the monopole, quadrupole, and hexadecapole terms of $P^s(k,\mu)$. The two measurements of the power spectrum agree well at large scales. At the small scales represented by the outermost contours, non-linearity has introduced power into the higher harmonics of $P^s(k,\mu)$, so our three-term reconstruction does not yield a complete description. By including more terms of the harmonic expansion, we could presumably recover $P^s(k,\mu)$ more accurately in this regime.

We do not need a complete reconstruction of $P^s(k,\mu)$ to apply our $\beta$ estimator (2.4), just the monopole and quadrupole moments. The solid line of Figure 7 shows the value of $\beta$ estimated from the direct FFT measurement of $P^s(k,\mu)$, as in Figure 2a. Open circles show the estimates of $\beta$ based on the monopole and quadrupole moments recovered by our "practical" method. The two estimates agree well at all wavelengths. However, the estimated $\beta$'s lie below the true value, especially at smaller wavelengths, because of the non-linear effects discussed in §2.2, with dispersion non-linearity being the most serious of these. To obtain accurate estimates of $\beta$ from a real redshift survey, one must either correct for non-linearities or measure the power spectrum on still larger scales.

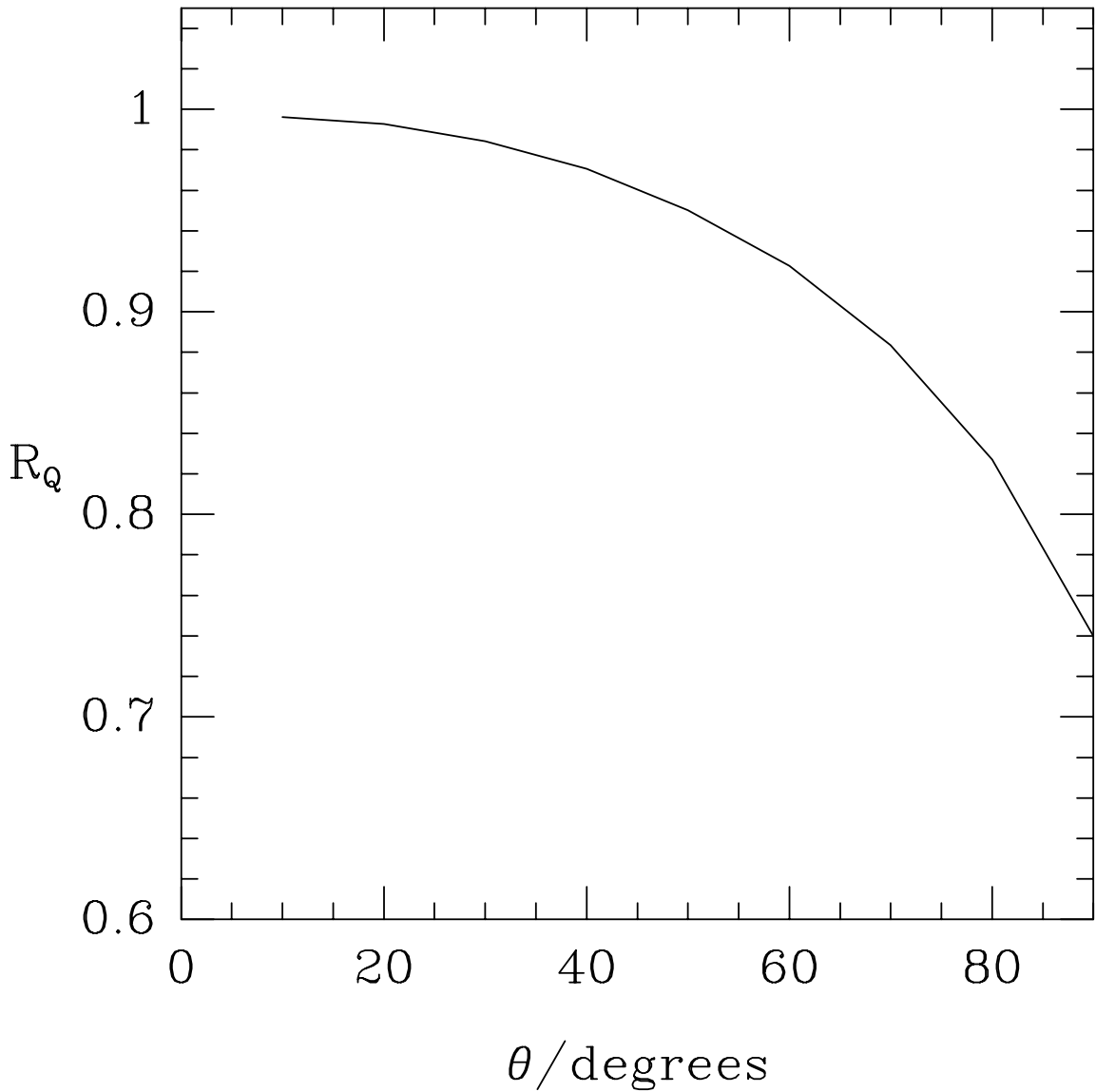

**Figure 8.** The ratio $R_Q$ of the quadrupole measured when the windowed data subtend a finite opening angle $\theta$ to the value obtained for a vanishingly small opening angle. This estimate was made for a wavelength $\lambda = 80\mathrm{Mpc/h}$ in an $\Omega = 1$, $b = 2$ $N$-body simulation. However, the effect is purely geometrical, and therefore independent of wavelength or the intrinsic value of $\beta$. Corrections for the finite opening angle must be applied when $\theta$ exceeds $\sim 30$ degrees.

In a real redshift survey we cannot place our window at infinite distance, so the data sample must subtend a finite opening angle. When the opening angle is large, the power spectrum estimator (3.3) averages over a range of line-of-sight directions $\mu_{\mathbf{kl}}$ at fixed $\mathbf{k}$, which depresses the amplitude of the measured quadrupole moment and thereby reduces the estimated value of $\beta$. We can use the $N$-body simulations to examine the magnitude of this effect. To do so, we measure $P^s(k,\mu)$ with our bowler-hat window function, but we set the observer at a distance $L = R_{sph}/\tan(\theta/2)$ from the window centre. Figure 8 shows the ratio $R_Q$ of the quadrupole measured with opening angle $\theta$ to the asymptotic value obtained with a vanishingly small opening angle, for $0 < \theta < 90°$. We see that $R_Q > 0.98$ for $\theta \lesssim 30°$, i.e. the measured quadrupole is diluted by less than 2% and the resulting estimate of $\beta$ is nearly unbiased. The dilution effect increases rapidly at larger angles. We have used a wavelength $\lambda = 80\mathrm{h}^{-1}\mathrm{Mpc}$ to compute Figure 8, but the dilution is a purely geometrical effect reflecting the spread in $\mu_{\bf kl}$ values within the window, so the quantity $R_Q$ should be independent of wavelength and of the intrinsic value of $\beta$. When analyzing a real redshift survey we can either ensure that the opening angle is less than 30°, or, if the limited survey depth requires that we use a larger opening angle, we can make a systematic correction to the measured quadrupole based on the ratio $R_Q$.

Our algorithm for measuring the redshift space power spectrum and estimating $\beta$ can be summarized as follows:

(1) For each wavenumber $k$, pick a random line of sight **l** and place a bowler-hat window at a distance $L \geq R_{sph}/\tan(\theta_{max}/2)$, so that it subtends an opening angle $\theta \leq \theta_{max}$. The window radius $R_{sph} = 4.4934/k$ is chosen to make the Fourier transform $\tilde{w}(k) = 0$.

(2) For many random choices of the wavevector direction $\hat{\bf k}$, carry out the Fourier sums (3.1) over galaxies in the window to estimate the mode amplitudes $S_{\bf k}$. Compute $\langle S_{\bf k} S_{\bf k}^* \rangle$, the squared-moduli averaged in bins of $\mu_{\bf kl}$, and subtract the shot noise term of equation (3.3) to obtain an estimate of $P^{sc}(k, \mu_{\bf kl})$, the convolved redshift space power spectrum.

(3) Increase the signal-to-noise ratio by repeating this procedure for many random choices of the window centre, within the volume defined by the survey boundaries and the opening angle constraint, and averaging the results.

(4) Fit the multipole moments $P_l^{sc}(k)$ of the measured $P^{sc}(k, \mu_{\bf kl})$. Correct these using the transfer functions $T_l^C(k)$, defined by equation (3.5), to obtain the moments $P_l^s(k)$ of the true redshift space power spectrum. If the maximum opening angle $\theta_{max}$ exceeds 30°, divide the quadrupole moment by the geometrical correction factor $R_Q$ plotted in Figure 8, and apply similar corrections for any higher moments of interest.

At this point, one can reconstruct $P^s(k, \mu)$ from its estimated multipole moments $P_l^s(k)$, and one can use the ratio $P_2^s(k)/P_0^s(k)$ to estimate $\beta$ via equation (2.4). The $\beta$ estimator is valid only in the linear regime, but the technique for measuring $P^s(k, \mu)$ is general and applies on all scales.

## 4 Preliminary Application to the IRAS 1.2 Jy Survey

We now present a preliminary application of our method to the 1.2 Jy IRAS survey studied by FDSYH. The sample we analyse is complete to a flux limit of 1.2 Jy at 60 $\mu$m and contains 5304 galaxies with galactic latitude $|b| > 5°$. A detailed description of the sample and its selection criteria can be found in Strauss et al. (1990) and Fisher (1992).

We use the method described in §3 to measure the angular dependence of the redshift space power spectrum, $P^s(k, \mu)$, at a variety of wavelengths. For each wavelength $\lambda \equiv 2\pi/k$, we use bowler-hat data windows [equation (3.2)], with $R_{sph} = 4.4934/k$ and $R_g = 0.1 R_{sph}$. We centre each window at a random angular position on a shell of radius $L = R_{sph}/\tan(55°/2)$, avoiding any overlap with the sample boundary at $|b| = 5°$. The longer the wavelength the more distant the shell, and therefore the larger the contribution of shot noise to the measured power. We adopt the rather large opening angle of $\theta = 55°$ in order to keep the shot noise tolerably small. The large opening angle slightly reduces the amplitude of the measured quadrupole component of the power spectrum, as described in §3. We correct this geometrical effect using the appropriate value of $R_Q$ read from Figure 8. In order to sample the data on each shell completely, we take 500 sub-samples defined by different, random angular positions of the bowler-hat window. In each sample we estimate the power for 50 randomly oriented **k**-vectors, using the selection function $\phi(s)$ adopted by FDSYH. We average the resulting estimates as a function of $\mu_{\bf kl}$, in bins of width $\Delta\mu_{\bf kl} = 0.1$.

At wavelengths $\lambda = 25, 30, 35$ and $40 h^{-1}$Mpc, the estimated power spectrum shows a clear enhancement for wavevectors close to the line of sight. The estimate of $P^s(k, \mu)$ at $\lambda = 50 h^{-1}$Mpc is considerably noisier than at the shorter wavelengths. This limitation is inevitable with the present data set, since at $50 h^{-1}$Mpc the shot noise term that is subtracted from the initial power estimate is as large as the remaining power. Figure 9 plots the values of $\beta$ deduced from the ratios of the quadrupole to monopole components at each of these wavelengths, after applying the transfer function correction appropriate for spectral index $n = -1$ (from Table 1) and the 4% geometrical correction for the 55° opening angle (from Figure 8). In the range $30\,h^{-1}\,\text{Mpc} < \lambda < 45\,h^{-1}\,\text{Mpc}$, the estimates are consistent with $\beta = 0.35 \pm 0.05$. However, our experience with the $N$-body simulations shows that the effect of velocity dispersions in groups and clusters is not negligible on these scales, and that this value is likely to be an underestimate.

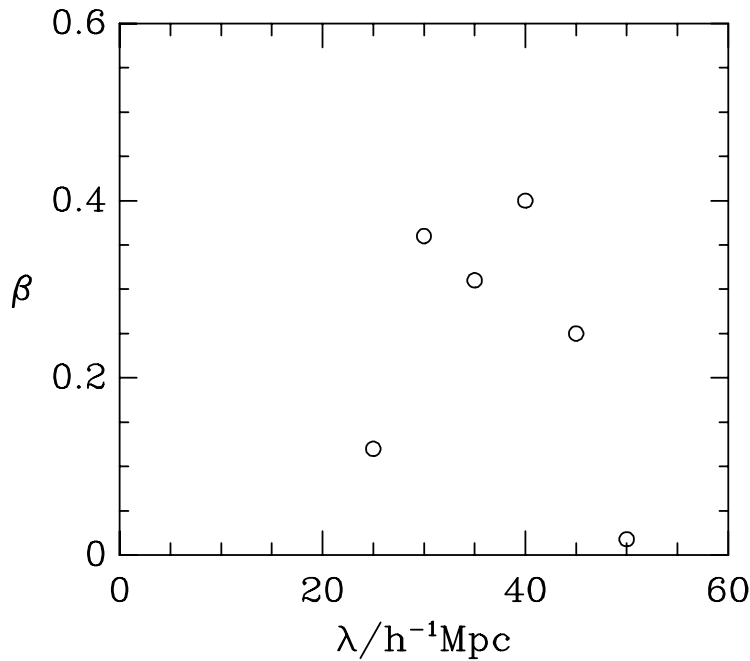

**Figure 9.** Values of $\beta$ estimated from the redshift space power spectrum of the 1.2 Jy *IRAS* redshift survey, using the linear theory estimator (2.4), as a function of wavelength $\lambda \equiv 2\pi/k$. Results for $30\,h^{-1}\,\text{Mpc} < \lambda < 45\,h^{-1}\,\text{Mpc}$ are consistent with $\beta = 0.35 \pm 0.05$. Non-linear effects remain important on these scales, so the true value of $\beta$ is probably higher. Errors due to shot noise become large beyond $\lambda \sim 45\,h^{-1}\,\text{Mpc}$.

This analysis of the 1.2 Jy *IRAS* survey is preliminary in several senses. Most obvious is the fact that Figure 9 contains no error bars. An estimate of the errors under the assumption of Gaussian fluctuations could be carried out using the formalism of Feldman *et al.* (1993), but a more thorough error analysis would require the use of mock *IRAS* catalogues constructed from $N$-body simulations, along the lines of FDSYH. We plan to perform such an analysis in a future, detailed study of the 1.2 Jy survey. Coupled to the absence of error bars is our decision to use, at each wavenumber $k$, only that portion of the survey that lies on a shell defined by the opening angle $\theta = 55°$. Utilizing the whole survey would require us to combine estimates from different shells with different shot noise, and taking a proper weighted average of the results would require knowledge of the errors on individual estimates. Because the shot noise increases rapidly with distance, it is not clear that we would gain much by including more distant shells in our power spectrum estimates, but a modest improvement in signal-to-noise should be possible. Finally we have not made any correction for non-linear effects, in particular for the effects of small scale velocity dispersions. We plan to use the simulated *IRAS* catalogues mentioned above to investigate possible methods of supressing dispersion non-linearity, such as compressing clusters or smoothing the data in redshift space. By combining results from different wavelengths, utilizing the survey volume in a more optimal manner, and correcting for dispersion non-linearity, we should be able to improve our estimate of $\beta$ significantly.

## 5 Conclusions

By shifting galaxies along the line of sight, peculiar motions make the redshift space power spectrum of galaxy clustering, $P^s(k,\mu)$, anisotropic. In linear perturbation theory, the degree of anisotropy depends only on the parameter $\beta \approx \Omega^{0.6}/b$, and the form is characterized entirely by its monopole, quadrupole, and hexadecapole moments. One can estimate the value of $\beta$ from the ratio of the quadrupole and monopole moments of $P^s(k,\mu)$, for any wavenumber $k$ in the linear regime.

We have described and tested a practical method of measuring $P^s(k,\mu)$ from galaxy redshift surveys, which uses an offset data window to select galaxies that lie close to the same line of sight. This method applies in both the linear and non-linear regimes. Key features of the method are a correction for discreteness contributions to the power spectrum and a correction for the systematic effect of selecting galaxies within a finite region. There are many advantages to analyzing redshift space distortions using the power spectrum. The prediction of linear theory takes a very simple form in Fourier space. Estimates of the power at different wavenumbers are statistically independent, provided they are separated by more than the width of the window in Fourier space, so estimates of $\beta$ at different wavelengths can be combined in a straightforward way to yield a more accurate overall estimate. Also, estimates of the power spectrum are insensitive to errors in the mean density of a survey, unlike estimates of the correlation function on large scales.

We have used a series of $N$-body simulations to explore the effects of non-linear gravitational evolution on the power spectrum anisotropy. Non-linearities take a variety of forms, and they can influence the power spectrum out to wavelengths of $50h^{-1}$Mpc or more. These effects complicate the task of inferring $\beta$ from $P^s(k,\mu)$ — typically, they cause our quadrupole method to underestimate the true value of $\beta$. However, the transition between the linear and non-linear regimes depends separately on $\Omega$ and $b$, so with sufficiently good data it should be possible to constrain both parameters individually.

The present study leaves at least three challenging tasks for future work. The first is to devise a practical method for removing or minimizing the effects of small scale velocity dispersions, since these seem to be the most pernicious and least useful form of non-linearity. The second is to calculate the uncertainties and correlations of power spectrum estimates and to use the results of this calculation to develop an optimal strategy for analyzing flux-limited data sets and combining estimates of $\beta$ from different scales. The third is to obtain a more thorough understanding of "dynamical" non-linearity — the breakdown of the linear theory relation between velocity and density perturbations — since this form of non-linearity is the most useful for separating $\Omega$ and $b$.

Our preliminary analysis of the 1.2 Jy *IRAS* redshift survey reveals a clear signal of anisotropy in the power spectrum on large scales. The quadrupole-to-monopole ratio yields $\beta_{est} = 0.38, 0.32$, and 0.42 at wavelengths $\lambda = 30, 35,$ and $40h^{-1}$Mpc, respectively. The residual effects of small scale velocity dispersions remain important at these wavelengths, so we expect these values to underestimate the true value of $\beta$. The limited depth of the 1.2 Jy survey prevents us from making reliable estimates at significantly larger scales, but suitable corrections for non-linearity should allow us to make the estimate from these wavelengths more accurate.

Prospects for applying this technique to other data sets are encouraging. The sparsely sampled QDOT survey of *IRAS* galaxies goes to a fainter flux limit than the 1.2 Jy survey, and it will soon be extended to a complete survey of *IRAS* galaxies brighter than 0.6 Jy. On the optical side, the extension of the CfA survey is nearly complete, extension of the Southern Sky Redshift Survey is underway, and the multi-fibre survey of Kirshner, Oemler, Schechter & Schectman is progressing rapidly. All of these data sets should provide useful, and largely independent, estimates of $\beta$. The million-galaxy redshift survey of the Sloane Digital Sky Survey should yield a precise measurement of $\beta$ and separate constraints on $\Omega$ and $b$.

**Acknowledgements**

SMC thanks Andy Jackson for acquainting him with some of the more obscure properties of Legendre polynomials which facilitated much of this work. We are grateful to Michael Strauss for helpful comments on the manuscript and to Changbom Park for the use of his $N$-body code. We thank Marc Davis, Michael Strauss, Amos Yahill and John Huchra for kindly allowing us access to the 1.2 Jy *IRAS* survey prior to publication. SMC and KBF acknowledge SERC postdoctoral fellowships. DHW acknowledges the support of a W.M. Keck Foundation fellowship and additional support from the Ambrose Monell Foundation and NSF grant PHY92-45317.

**Appendix A. The Convolved Redshift Space Power Spectrum**

Suppose that we have a redshift survey characterized by a mean galaxy density $\langle n \rangle$ and radial selection function $\phi(r)$, defined so that the number of galaxies per unit volume at distance $r$ would be $\langle n \rangle \phi(r)$ in the absence of inhomogeneities. As an estimator for the Fourier amplitudes, consider the quantity

$$S_{\mathbf{k}} = \frac{1}{\langle n \rangle V_w^{\frac{1}{2}}} \sum_{i=1}^{N} \frac{w(\mathbf{s}_i)}{\phi(\mathbf{s}_i)} e^{+i\mathbf{k}\cdot\mathbf{s}_i} \ , \tag{A.1}$$

where $N$ is the total number of galaxies in the survey and $V_w \equiv \int d^3\mathbf{s}\, w(\mathbf{s})$. The window function $w(\mathbf{s})$ defines the limits of the data sample used to compute $S_{\mathbf{k}}$, since data lying in the region where $w(\mathbf{s}) = 0$ do not contribute to the Fourier sum. However, $w(\mathbf{s})$ need not be constant over its non-zero region, and it need not be centred on the observer. In this paper, for example, we adopt the bowler-hat window described in §3, and we centre it far enough from the observer that it subtends an opening angle $\theta \leq 55°$.

Following Peebles (1980, §41.1), we rewrite $S_{\mathbf{k}}$ as a sum over infinitesimal cells with volumes $d^3\mathbf{s}_i$ and occupation numbers $N_i$:

$$S_{\mathbf{k}} = \frac{1}{\langle n \rangle V_w^{\frac{1}{2}}} \sum_{cells\ i} \frac{N_i w(\mathbf{s}_i)}{\phi(\mathbf{s}_i)} e^{+i\mathbf{k}\cdot\mathbf{s}_i} \ . \tag{A.2}$$

The ensemble average of the squared-modulus of $S_{\mathbf{k}}$ is

$$\langle S_{\mathbf{k}} S_{\mathbf{k}}^* \rangle = \frac{1}{\langle n \rangle^2 V_w} \sum_{cells\ i,j} \langle N_i N_j \rangle \frac{w(\mathbf{s}_i) w^*(\mathbf{s}_j)}{\phi(\mathbf{s}_i)\phi(\mathbf{s}_j)} e^{+i\mathbf{k}\cdot(\mathbf{s}_i - \mathbf{s}_j)} \ . \tag{A.3}$$

The expected occupation numbers of cells located at positions $\mathbf{s}_i$ and $\mathbf{s}_j$ are

$$\langle N_i N_j \rangle = \begin{cases} \langle n \rangle^2 \phi(s_i)\phi(s_j) d^3\mathbf{s}_i\, d^3\mathbf{s}_j \left( 1 + \langle \delta^s(\mathbf{s}_i)\delta^{s*}(\mathbf{s}_j) \rangle \right), & i \neq j \\ \langle n \rangle \phi(s_i) d^3\mathbf{s}_i, & i = j \end{cases} \ . \tag{A.4}$$

For the $i = j$ term, we have used the fact that $N_i$ must equal 0 or 1 in infinitesimal cells, implying $N_i^2 = N_i$ and $\langle N_i^2 \rangle = \langle N_i \rangle$. In real space $\langle \delta^R(\mathbf{x}_i)\delta^{R*}(\mathbf{x}_j) \rangle$ depends only on the distance $x = |\mathbf{x}_i - \mathbf{x}_j|$ and is given by the usual two-point correlation function, $\xi^R(x)$. In redshift space the correlation function, like the power spectrum, is anisotropic (Kaiser 1987; Lilje & Efstathiou 1989; M$^c$Gill 1990; Hamilton 1992), and it depends both on the length of the separation vector, $s = |\mathbf{s}_i - \mathbf{s}_j|$, and on the vector's orientation with respect to the line of sight. The correlation function $\xi^s(s, \mu_{\mathbf{sl}})$ and the power spectrum $P^s(k', \mu_{\mathbf{k'l}})$ form a Fourier transform conjugate pair, with

$$\langle \delta^s(\mathbf{s}_i)\delta^{s*}(\mathbf{s}_j) \rangle \equiv \xi^s(s, \mu_{\mathbf{sl}}) = \frac{1}{(2\pi)^3} \int d^3\mathbf{k}'\, P^s(k', \mu_{\mathbf{k'l}}) e^{-i\mathbf{k'}\cdot\mathbf{s}} \ , \tag{A.5}$$

where $\mathbf{s} = \mathbf{s}_i - \mathbf{s}_j$. Substituting equation (A.5) into equation (A.4) allows equation (A.3) to be written, after some manipulation, as

$$\langle S_{\mathbf{k}} S_{\mathbf{k}}^* \rangle = \frac{V_w}{(2\pi)^3} \int d^3\mathbf{k}'\, P^s(k', \mu_{\mathbf{k'l}}) |\tilde{w}(\mathbf{k} - \mathbf{k}')|^2 + V_w |\tilde{w}(\mathbf{k})|^2 + \frac{1}{\langle n \rangle V_w} \int d^3\mathbf{s}\, \frac{|w(\mathbf{s})|^2}{\phi(\mathbf{s})} \ , \tag{A.6}$$

where $\tilde{w}(\mathbf{k}) \equiv V_w^{-1} \int w(\mathbf{s})e^{+i\mathbf{k}\cdot\mathbf{s}}d^3\mathbf{s}$ is the Fourier transform of the window function, normalized to unity at $\mathbf{k} = 0$.

The first term in equation (A.6) is the true redshift space power spectrum convolved with the squared Fourier transform of the data window. The second term arises from the finite width of the window function in Fourier space; in the limit of a large volume, this term would contribute only at $\mathbf{k} = \mathbf{0}$. The last term is the discreteness or "shot noise" contribution to the Fourier transform.

When estimating the power spectrum for a given value of $k$, we choose the radius of the data window to yield $\tilde{w}(\mathbf{k}) = 0$ (following FDSYH). We then compute the squared-modulus of the Fourier sum defined by equation (A.1) and subtract the shot noise (final) term of equation (A.6). The line of sight is the only preferred direction, so we can average results over different directions of $\mathbf{k}$ that have the same $\mu_{\mathbf{kl}}$, and we can average over different positions of the data window within the redshift survey. We are left with an estimate of the convolved redshift space power spectrum,

$$P^{SC}(k, \mu_{\mathbf{kl}}) \equiv \frac{V_w}{(2\pi)^3} \int d^3\mathbf{k}' \, P^S(k', \mu_{\mathbf{k'l}}) \, |\tilde{w}(\mathbf{k} - \mathbf{k}')|^2 \ . \tag{A.7}$$

The effect of the convolution on $P^S(k, \mu)$ can be conveniently expressed in terms of multipole moments. We proceed by first decomposing the true and convolved power spectra and the window transform factor $|\tilde{w}(\mathbf{k} - \mathbf{k}')|^2$ into their multipole moments,

$$P^S(k, \mu) = \sum_{l=0}^{\infty} P_l^S(k)\mathcal{P}_l(\mu) \ , \quad P^{SC}(k, \mu) = \sum_{l=0}^{\infty} P_l^{SC}(k)\mathcal{P}_l(\mu) \ ,$$

$$\text{and} \quad |\tilde{w}(\mathbf{k} - \mathbf{k}')|^2 = \sum_{l=0}^{\infty} \tilde{w}_l(k, k')\mathcal{P}_l(\mu_{\mathbf{kk'}}) \ . \tag{A.8}$$

The expansion coefficients can be obtained from the inversions

$$P_l^S(k) \equiv \frac{2l+1}{2} \int_{-1}^{+1} d\mu \, P^S(k, \mu)\mathcal{P}_l(\mu) \ , \quad P_l^{SC}(k) \equiv \frac{2l+1}{2} \int_{-1}^{+1} d\mu \, P^{SC}(k, \mu)\mathcal{P}_l(\mu) \ ,$$

$$\tilde{w}_l(k, k') \equiv \frac{2l+1}{2} \int_{-1}^{+1} d\mu_{\mathbf{kk'}} \, |\tilde{w}(\mathbf{k} - \mathbf{k}')|^2 \, \mathcal{P}_l(\mu_{\mathbf{kk'}}) \ . \tag{A.9}$$

In equations (A.8) and (A.9), $\mathcal{P}_l(\mu)$ denotes the Legendre polynomial of order $l$.

With the expansions given in (A.8), the convolved power spectrum appearing in (A.7) can be written as

$$P^{SC}(k, \mu_{\mathbf{kl}}) = \sum_{l'} \sum_{l} \frac{V_w}{(2\pi)^3} \int_0^{\infty} dk' \, k'^2 \, P_l^S(k')\tilde{w}_{l'}(k, k') \oint d\Omega_{\mathbf{k}'} \, \mathcal{P}_l(\mu_{\mathbf{k'l}})\mathcal{P}_{l'}(\mu_{\mathbf{kk'}}) \ . \tag{A.10}$$

We can carry out the sum over $l'$ by applying the following theorem for Legendre polynomials,

$$\oint d\Omega_{\mathbf{k}'} \, \mathcal{P}_l(\mu_{\mathbf{k's_1}})\mathcal{P}_{l'}(\mu_{\mathbf{k's_2}}) = \frac{4\pi}{2l+1}\mathcal{P}_l(\mu_{\mathbf{s_1 s_2}})\delta_{ll'}^K \ , \tag{A.11}$$

where $\delta_{ll'}^K$ is the Kronecker delta symbol. Theorem (A.11) reduces equation (A.10) to

$$P^{SC}(k, \mu_{\mathbf{kl}}) = \sum_l \left[ \frac{4\pi}{2l+1} \frac{V_w}{(2\pi)^3} \int_0^{\infty} dk' \, k'^2 \, P_l^S(k')\tilde{w}_l(k, k') \right] \mathcal{P}_l(\mu_{\mathbf{kl}}) \ . \tag{A.12}$$

Recalling the definition of $\tilde{w}_l(k, k')$, this becomes

$$P^{sc}(k, \mu_{\mathbf{kl}}) = \sum_l \left[ \frac{V_w}{(2\pi)^2} \int_0^\infty dk'\, k'^2 P_l^s(k') \int_{-1}^{+1} d\mu_{\mathbf{kk'}}\, |\tilde{w}(\mathbf{k} - \mathbf{k'})|^2 \mathcal{P}_l(\mu_{\mathbf{kk'}}) \right] \mathcal{P}_l(\mu_{\mathbf{kl}}) \,. \qquad (A.13)$$

By comparing (A.13) with the expansion for $P^{sc}(k, \mu)$ given in (A.8), we see that the terms in the square brackets are just the multipole moments of the convolved spectrum, $P_l^{sc}(k)$, and that these are related to the unconvolved coefficients by the following transfer function,

$$\begin{aligned} T_l^C(k) &\equiv \frac{P_l^{sc}(k)}{P_l^s(k)} \\ &= \frac{1}{P_l^s(k)} \frac{V_w}{(2\pi)^2} \int_0^\infty dk'\, k'^2 P_l^s(k') \int_{-1}^{+1} d\mu_{\mathbf{kk'}}\, |\tilde{w}(\mathbf{k} - \mathbf{k'})|^2 \mathcal{P}_l(\mu_{\mathbf{kk'}}) \\ &= \frac{1}{P_l^s(k)} \int_0^\infty dk'\, k'^2 P_l^s(k') \mathcal{F}_l(k, k') \,, \end{aligned} \qquad (A.14)$$

where the convolution kernel appearing in the last line is

$$\mathcal{F}_l(k, k') = \frac{V_w}{(2\pi)^2} \int_{-1}^{+1} d\mu_{\mathbf{kk'}}\, |\tilde{w}(\mathbf{k} - \mathbf{k'})|^2 \mathcal{P}_l(\mu_{\mathbf{kk'}}) \,. \qquad (A.15)$$

Note that $\mathcal{F}_l(k, k')$ depends only on the window geometry, not on the power spectrum. Furthermore, for a sensible choice of window $\mathcal{F}_l(k, k')$ is sharply peaked around $k = k'$, so the power spectrum factors inside and outside the integral of equation (A.14) nearly cancel, making the transfer function insensitive to the shape of $P_l^s(k)$ (cf. Table 1). In the limit that the window function $w(\mathbf{s})$ encompasses a large volume, the effect of the convolution become negligible, $\mathcal{F}_l(k, k') \to k^{-2} \delta^{(1)}(k - k')$, and we recover $T_l^C(k) \equiv 1$.

### Appendix B. The Redshift Space Correlation Function

In this Appendix we derive identities relating the harmonic components of the redshift space correlation function to those of the redshift space power spectrum. We use these identities to show that in the linear regime our formalism reproduces the expressions derived by Hamilton (1992).

The redshift space correlation function and power spectrum form a Fourier conjugate pair,

$$\xi^s(s, \mu_{\mathbf{sl}}) = \frac{1}{(2\pi)^3} \int d^3\mathbf{k}\, P^s(k, \mu_{\mathbf{kl}}) e^{-iks\mu_{\mathbf{ks}}} \,. \qquad (B.1)$$

The vector $\mathbf{l}$ denotes the line of sight direction, and $\mu_{\mathbf{ks}}$ denotes the cosine of the angle between the vectors $\mathbf{s}$ and $\mathbf{k}$. We now replace $P^s(k, \mu)$ by its expansion in terms of Legendre polynomials,

$$P^s(k, \mu_{\mathbf{kl}}) = \sum_l P_l^s(k)\, \mathcal{P}_l(\mu_{\mathbf{kl}}) \,, \qquad (B.2)$$

and we substitute for $e^{-iks\mu_{\mathbf{ks}}}$ using the Rayleigh expansion of a plane wave,

$$e^{-iks\mu_{\mathbf{ks}}} = \sum_l (-i)^l (2l+1) j_l(ks) \mathcal{P}_l(\mu_{\mathbf{ks}}) \,, \qquad (B.3)$$

where $j_l(ks)$ is the spherical Bessel function. We obtain

$$\xi_l^s(s, \mu_{\mathbf{sl}}) = \sum_l \sum_{l'} (-i)^{l'} (2l'+1) \frac{1}{(2\pi)^3} \int_0^\infty dk\, k^2 P_l^s(k) j_{l'}(ks) \oint d\Omega_{\mathbf{k}} \mathcal{P}_l(\mu_{\mathbf{kl}}) \mathcal{P}_{l'}(\mu_{\mathbf{ks}}) \,. \qquad (B.4)$$

Applying theorem (A.11) yields the relation

$$\xi_l^s(s) = i^l \frac{1}{2\pi^2} \int_0^\infty dk \, k^2 P_l^s(k) j_l(ks) \,, \tag{B.5}$$

where $\xi_l^s(s)$ is the order-$l$ harmonic coefficient of the redshift space correlation function, i.e.

$$\xi^s(s, \mu_{\mathbf{sl}}) = \sum_l \xi_l^s(s) \, \mathcal{P}(\mu_{\mathbf{sl}}). \tag{B.6}$$

The odd-$l$ components of $P^s(k, \mu)$ and $\xi^s(s, \mu)$ vanish by symmetry, so the factor $i^l$ is simply $\pm 1$ and determines the relative sign of harmonic distortions in the correlation function and the power spectrum. The expansion defined by (B.6) and (B.5) is an identity, valid in both the linear and non-linear regime. In the case of linear theory, one can substitute for $P_l^s(k)$ using the relations (2.3). The resulting harmonic expansion of $\xi^s(s, \mu)$ is identical to the expression given by Fisher et al. (1993c).

In order to compare the Fourier analysis to the correlation function analysis presented by Hamilton (1992), the following identities are useful:

$$\frac{1}{2\pi^2} \int_0^\infty dk \, k^2 P_0^s(k) j_0(ks) = \xi_0^s(s) \,, \tag{B.7}$$

$$\frac{1}{2\pi^2} \int_0^\infty dk \, k^2 P_0^s(k) j_2(ks) = -\xi_0^s(s) + \frac{3}{s^3} \int_0^s ds' \, s'^2 \xi_0^s(s') \,, \tag{B.8}$$

$$\frac{1}{2\pi^2} \int_0^\infty dk \, k^2 P_0^s(k) j_4(ks) = \xi_0^s(s) + \frac{5}{2} \frac{3}{s^3} \int_0^s ds' \, s'^2 \xi_0^s(s') - \frac{7}{2} \frac{5}{s^5} \int_0^s ds' \, s'^4 \xi_0^s(s') \,. \tag{B.9}$$

The identity (B.7) is simply (B.5) for the case of $l = 0$, and the other two identities can be derived from (B.7) using the recurrence relations and integral properties of spherical Bessel functions. If one then substitutes the linear theory relations (2.3), these relations give equations (6), (7) and (8) of Hamilton (1992).